\documentclass[11pt]{article}

\PassOptionsToPackage{hyphens}{url}
\usepackage[utf8]{inputenc}
\usepackage[a4paper,margin=1in]{geometry}
\usepackage{amsmath,amssymb}
\usepackage{microtype}
\usepackage{booktabs}
\usepackage{tabularx}
\usepackage{array}
\usepackage{enumitem}
\usepackage[table]{xcolor}
\usepackage{listings}
\usepackage{tikz}
\usetikzlibrary{arrows.meta,positioning,fit,calc}
\usepackage{underscore}
\usepackage[hidelinks]{hyperref}
\usepackage{url}

\setlist{itemsep=2pt,topsep=4pt}
\newcolumntype{L}{>{\raggedright\arraybackslash}X}
\newcommand{\code}[1]{\texttt{#1}}
\newcommand{\XIM}{XIM}

\lstdefinestyle{spec}{
  basicstyle=\ttfamily\footnotesize,
  columns=fullflexible,
  keepspaces=true,
  frame=single,
  framerule=0.4pt,
  rulecolor=\color{black!40},
  backgroundcolor=\color{black!3},
  xleftmargin=4pt,xrightmargin=4pt,
  aboveskip=8pt,belowskip=8pt,
  breaklines=true,
  showstringspaces=false,
  upquote=false
}
\tikzset{
  st/.style={draw,rounded corners=2pt,minimum height=6mm,inner sep=3pt,font=\ttfamily\scriptsize,fill=black!4},
  bx/.style={draw,rounded corners=2pt,minimum height=8mm,minimum width=24mm,font=\small,fill=black!4,align=center},
  arr/.style={-{Stealth[length=2mm]},thick}
}

\title{\textbf{XIM: The XDC Interledger Messaging Protocol}\\[4pt]
\large A Verification-Agnostic Messaging and Settlement Fabric for Heterogeneous Ledgers and Financial Rails}

\author{Atul Khekade \qquad Ritesh Kakkad \qquad Wanwiset Peerapatanapokin \\[2pt] Behnam Mohammadkhani\\[4pt]
\normalsize XDC Network Research and Engineering}

\date{September 2026}

\begin{document}
\maketitle

\begin{abstract}
Distributed ledgers, privacy-preserving institutional networks, and conventional payment systems increasingly need to exchange authenticated messages and settle assets across heterogeneous trust domains. Existing interoperability systems typically optimize for one of three concerns: application-level abstraction, cross-chain message transport, or synchronized execution within a related ledger family. This paper proposes \XIM{}, the XDC Interledger Messaging Protocol, a chain-agnostic protocol for transporting canonical messages across heterogeneous networks while allowing each communication lane to select an explicit verification policy. \XIM{} separates message semantics from transport, verification, execution, routing, asset identity, and compliance metadata. Its cryptographic state is represented by deterministic message identifiers and commitment roots, while an append-only transition log provides auditability and replay protection. \XIM{} introduces a Universal Asset Identifier (UAID), a pluggable adapter interface, lane-scoped security policies, and an optional policy-aware route graph for multi-hop settlement. XDC Network can serve as a coordination and settlement domain without requiring every \XIM{} message or route to transact through XDC. We specify the protocol model, state machine, message encoding, commitment structure, verification modes, failure semantics, security assumptions, threat model, implementation architecture, and an incremental deployment plan. The design targets public blockchains, permissioned ledgers, institutional networks, and authenticated financial-system gateways, with particular attention to stablecoins, tokenized assets, trade finance, and ISO~20022-compatible payment workflows.
\end{abstract}

\noindent\textbf{Keywords:} blockchain interoperability; enterprise middleware; cross-chain messaging; interledger; XDC Network; Merkle commitments; light clients; zero-knowledge proofs; stablecoins; tokenized assets; ISO~20022; institutional settlement; routing; distributed ledgers.

%=====================================================================
\section{Introduction}
%=====================================================================

Financial infrastructure is becoming multi-ledger rather than converging on a single execution environment. Public smart-contract networks~\cite{nakamoto2008,wood-yellow}, permissioned ledgers, privacy-preserving institutional networks, tokenized deposit platforms, stablecoin networks, and conventional payment rails have different execution models, finality assumptions, privacy constraints, and governance. Point-to-point bridges do not scale organizationally or technically as the number of networks grows, while a single universal consensus system would require heterogeneous networks to abandon their native trust and privacy models.

\XIM{} addresses this problem by treating interoperability as authenticated message exchange between independently governed state machines. The protocol defines a canonical message, deterministic identifiers, explicit lifecycle states, cryptographic commitments, replay protection, verification-policy negotiation, destination execution, and acknowledgements. It intentionally does not mandate one consensus algorithm or one cross-chain proof mechanism. Instead, each source--destination lane binds to a verification policy appropriate to the value, finality, and trust characteristics of the connected systems.

The central design principle is separation of concerns. Transport moves bytes; verification establishes that a source event satisfies a lane's trust policy; routing selects an eligible path; execution applies a verified action at a destination; and settlement concerns the transfer or transformation of economic value. A compromise in one component should not automatically grant authority over all other components.

\subsection{Contributions}
\begin{itemize}
  \item A canonical, deterministic interledger message envelope independent of source-chain transaction format.
  \item Lane-scoped verification policies supporting native proofs, light clients, zero-knowledge proofs, threshold attestations, trusted-hardware attestations, and hybrid verification.
  \item Commitment-oriented state based on message hashes and Merkle roots rather than global replication of foreign-chain state.
  \item An explicit cross-domain message state machine with acknowledgement, timeout, refund, and terminal failure semantics.
  \item A Universal Asset Identifier (UAID) separating economic asset identity from chain-specific token contract addresses.
  \item A pluggable network adapter interface for public chains, permissioned ledgers, institutional networks, and authenticated legacy gateways.
  \item An optional policy-aware route graph for multi-hop settlement based on cost, latency, liquidity, security, privacy, jurisdiction, and finality.
  \item Separation of routing and execution infrastructure from an independent lane risk-monitoring function capable of rate-limiting or pausing a compromised lane.
\end{itemize}

\subsection{Non-goals}
\begin{itemize}
  \item \XIM{} is not a new general-purpose Layer-1 blockchain.
  \item \XIM{} does not require global replication of all connected-ledger transactions.
  \item \XIM{} does not claim trustless interoperability where the connected source cannot produce independently verifiable evidence.
  \item \XIM{} does not define universal legal finality; legal settlement finality remains dependent on the relevant asset, institution, jurisdiction, and network rules.
  \item \XIM{} does not require XDC to be an intermediate asset or mandatory hop for every route.
\end{itemize}

\subsection{Conventions}
The key words MUST, MUST NOT, SHOULD, SHOULD NOT, and MAY are to be interpreted as described in BCP~14~\cite{rfc2119,rfc8174} when, and only when, they appear in all capitals. $H(\cdot)$ denotes a collision-resistant hash function and $\|$ denotes byte-string concatenation.

%=====================================================================
\section{Related Work and Design Context}
%=====================================================================

Interoperability systems can be viewed along several architectural axes. API-gateway approaches abstract network-specific interfaces behind standardized application APIs~\cite{quant-overledger,quant-options}. Cross-chain messaging protocols authenticate source events and cause destination execution~\cite{ccip}. Light-client protocols verify counterparty consensus state and packet commitments~\cite{ibc,goes2020ibc}. Privacy-preserving institutional networks coordinate selectively disclosed transactions among authorized participants~\cite{canton-protocol,canton2020,canton-wp}. \XIM{} adopts concepts from each family but defines a distinct protocol boundary: a canonical message and state-transition system with pluggable verification and execution. Table~\ref{tab:related} summarizes the lessons drawn from representative systems.

\begin{table}[!htbp]
\centering\small
\caption{Representative interoperability systems and the lessons \XIM{} draws from them.}
\label{tab:related}
\begin{tabularx}{\linewidth}{@{}>{\raggedright\arraybackslash}p{0.22\linewidth}LL@{}}
\toprule
\textbf{System / family} & \textbf{Primary abstraction} & \textbf{Relevant lesson for \XIM{}}\\
\midrule
Quant Overledger / API gateways~\cite{quant-overledger,quant-options} & Unified API over heterogeneous DLT and legacy systems & Application abstraction and connector modularity.\\
Chainlink CCIP~\cite{ccip} & Cross-chain messages and tokens with separate commit/execute security components & Separation of routing, verification, execution, and independent risk controls.\\
IBC~\cite{ibc,goes2020ibc} & Clients, connections/channels, packet commitments, acknowledgements, and timeouts & Explicit counterparty verification and packet lifecycle semantics.\\
Circle CCTP~\cite{cctp} & Issuer-mediated burn/mint with signed attestation & Prefer canonical native asset movement over wrapped representations when issuer support exists.\\
Canton~\cite{canton-protocol,canton2020,canton-wp} & Privacy-aware synchronization across participant nodes and applications & Interoperability can preserve selective disclosure rather than forcing global data replication.\\
\bottomrule
\end{tabularx}
\end{table}

Beyond production systems, the academic literature provides systematizations of cross-ledger communication. Zamyatin et al.~\cite{zamyatin2021sok} show that correct cross-chain communication cannot be achieved without a trusted third party or additional assumptions, which motivates \XIM{}'s requirement that every lane name its trust assumption explicitly. Belchior et al.~\cite{belchior2021survey} survey the broader design space of blockchain interoperability. Herlihy's atomic cross-chain swaps~\cite{herlihy2018atomic} underpin the hash- and time-conditioned settlement discussed in Section~\ref{sec:atomicity}, and zkBridge~\cite{xie2022zkbridge} demonstrates practical succinct verification of source-chain consensus of the kind assumed by the \code{ZK\_PROOF} verification mode. Empirical analyses of cross-chain bridge exploits~\cite{lee2023bridges} motivate the lane-scoped limits and circuit breakers of Section~\ref{sec:risk}. Finally, the term \emph{interledger} is also associated with the Interledger Protocol (ILP)~\cite{thomas2015ilp}, which routes payments across ledgers through connectors using conditional transfers. \XIM{} differs in scope: it carries general authenticated messages rather than only payments, and it binds an explicit, versioned verification policy to each lane rather than relying on connector-held liquidity and hashlock conditions.

%=====================================================================
\section{System Model}
%=====================================================================

Let $\mathcal{N} = \{N_1, N_2, \ldots, N_n\}$ be a set of independently governed networks. A network may be a public blockchain, permissioned DLT, institutional ledger, or authenticated financial-system gateway. \XIM{} defines directed communication lanes $L_{ij}$ from source network $N_i$ to destination network $N_j$. A lane is identified by its source network, destination network, protocol version, adapter versions, and verification policy.

A \XIM{} deployment may include Observers, Verifiers, Routers, Executors, Risk Monitors, and Gateways (Table~\ref{tab:roles}). Roles may be combined in an early deployment but are logically distinct.

\begin{table}[!htbp]
\centering\small
\caption{Logical roles in a \XIM{} deployment.}
\label{tab:roles}
\begin{tabularx}{\linewidth}{@{}>{\raggedright\arraybackslash}p{0.2\linewidth}L@{}}
\toprule
\textbf{Role} & \textbf{Responsibility}\\
\midrule
Observer & Detects finalized or policy-eligible source events and constructs candidate \XIM{} messages.\\
Verifier & Evaluates source evidence against the lane verification policy.\\
Router & Selects direct or multi-hop paths that satisfy policy constraints.\\
Executor & Submits a verified destination action and reports the resulting destination transaction.\\
Risk Monitor & Independently detects anomalies and may throttle or pause a lane according to governance rules.\\
Gateway & Maps non-DLT financial messages or authenticated APIs into and out of \XIM{} semantics.\\
\bottomrule
\end{tabularx}
\end{table}

\subsection{Trust Domains}
\XIM{} does not collapse connected networks into a single trust domain. The source network remains responsible for its own consensus and finality. The destination remains responsible for destination execution. The lane verification policy defines the evidence required for the destination to accept a source assertion. Consequently, security claims MUST be expressed per lane, not globally.

%=====================================================================
\section{Network and Asset Identifiers}
%=====================================================================

\subsection{Network Identifier}
\begin{equation}
\mathsf{NetworkID} := H\big(\mathsf{namespace} \,\|\, \mathsf{canonical\_network\_name} \,\|\, \mathsf{genesis\_or\_domain\_identifier}\big)
\end{equation}
A NetworkID MUST be globally stable for a specific trust domain. Forked or re-genesis networks SHOULD receive a new identifier when their security identity changes materially.

\subsection{Universal Asset Identifier (UAID)}
A UAID identifies an economic asset independently from its representation on a particular ledger. A UAID record is governance-controlled metadata and MUST NOT imply that two representations are economically equivalent unless the registry's issuer and conversion policy establishes that equivalence.

\par\noindent\begin{minipage}{\linewidth}
\begin{lstlisting}
UAIDRecord {
  uaid
  issuer_id
  asset_type
  denomination
  jurisdiction
  representations[]
  issuance_policy_hash
  redemption_policy_hash
  compliance_policy_hash
}
\end{lstlisting}
\end{minipage}\par\smallskip

For example, native USDC on multiple supported networks may map to one issuer-level asset identity, while a third-party wrapped token SHOULD use a distinct representation classification unless redemption equivalence is explicitly verified.

%=====================================================================
\section{Canonical \XIM{} Message}
%=====================================================================

\XIM{} messages MUST use deterministic serialization. The reference implementation SHOULD use deterministic Protocol Buffers or an equivalent canonical binary encoding. JSON MAY be exposed at API boundaries but MUST NOT be used directly for consensus-critical hashing unless canonical JSON rules are specified.

\begin{lstlisting}
XIMMessage {
  uint32  version;
  bytes32 source_network;
  uint64  source_height;
  bytes   source_tx_id;
  bytes   sender;

  bytes32 destination_network;
  bytes   receiver;

  uint32  action_type;
  bytes32 asset_id;
  uint256 amount;
  bytes32 payload_hash;

  uint64  nonce;
  uint64  created_at;
  uint64  expiry;

  uint32  finality_policy;
  uint32  proof_type;
  bytes32 proof_commitment;

  bytes32 compliance_metadata_hash;
  bytes32 privacy_policy_hash;
  bytes32 route_policy_hash;
}
\end{lstlisting}

Fixed-width types such as \code{bytes32} and \code{uint256}, which have no native Protocol Buffers equivalent, are carried as \code{bytes} in the wire encoding (Appendix~\ref{app:proto}); a conforming decoder MUST reject values whose length differs from the declared width, and \code{uint256} values MUST be encoded as 32-byte big-endian integers.

\subsection{Message Identifier}
\begin{equation}
\mathsf{message\_id} = H\big(\mathsf{domain\_separator} \,\|\, \mathsf{canonical\_encode}(\mathsf{XIMMessage})\big)
\end{equation}
The domain separator MUST bind the protocol name and major version. A destination MUST reject a previously consumed \code{message\_id}. This provides protocol-level replay protection independent of destination-chain nonce behavior.

\subsection{Action Types}
Table~\ref{tab:actions} lists the action types defined by this version of the protocol.

\begin{table}[!htbp]
\centering\small
\caption{\XIM{} action types.}
\label{tab:actions}
\begin{tabularx}{\linewidth}{@{}>{\raggedright\arraybackslash}p{0.25\linewidth}L@{}}
\toprule
\textbf{Action} & \textbf{Meaning}\\
\midrule
\code{MESSAGE} & Deliver an application payload commitment.\\
\code{TRANSFER} & Transfer or release an asset representation.\\
\code{MINT\_BURN} & Coordinate canonical issuer-mediated burn/mint.\\
\code{CALL} & Invoke a destination contract or application endpoint.\\
\code{ATOMIC\_LEG} & Represent one leg of a multi-leg conditional settlement.\\
\code{ACK} & Acknowledge a terminal or intermediate state.\\
\code{CANCEL} / \code{REFUND} & Request policy-authorized cancellation or compensation.\\
\bottomrule
\end{tabularx}
\end{table}

%=====================================================================
\section{Message Lifecycle and State Machine}
\label{sec:lifecycle}
%=====================================================================

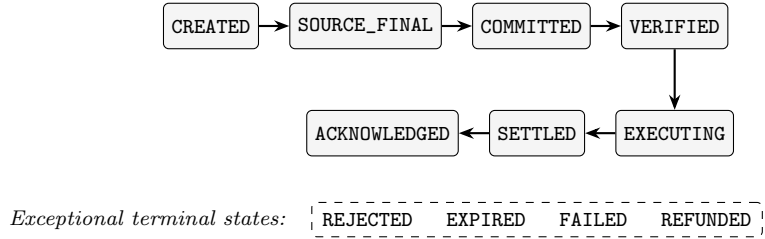
\begin{figure}[!htbp]
\centering
\begin{tikzpicture}[node distance=4mm]
  \node[st] (c) {CREATED};
  \node[st,right=of c] (sf) {SOURCE\_FINAL};
  \node[st,right=of sf] (cm) {COMMITTED};
  \node[st,right=of cm] (v) {VERIFIED};
  \node[st,below=8mm of v] (ex) {EXECUTING};
  \node[st,left=of ex] (se) {SETTLED};
  \node[st,left=of se] (ack) {ACKNOWLEDGED};
  \draw[arr] (c)--(sf); \draw[arr] (sf)--(cm); \draw[arr] (cm)--(v);
  \draw[arr] (v)--(ex); \draw[arr] (ex)--(se); \draw[arr] (se)--(ack);
  \node[draw,dashed,rounded corners=2pt,inner sep=4pt,font=\ttfamily\scriptsize,below=6mm of se] (term)
    {REJECTED \quad EXPIRED \quad FAILED \quad REFUNDED};
  \node[font=\scriptsize\itshape,left=2mm of term] {Exceptional terminal states:};
\end{tikzpicture}
\caption{Nominal \XIM{} message lifecycle (solid) and exceptional terminal states (dashed). The order along the solid path defines the lifecycle order $\prec$ used in Section~\ref{sec:invariants}.}
\label{fig:lifecycle}
\end{figure}

Figure~\ref{fig:lifecycle} shows the nominal lifecycle of a \XIM{} message together with its exceptional terminal states. State transitions MUST be monotonic except where an explicit compensating transaction creates a new message. \XIM{} SHOULD avoid pretending that irreversible execution can be rolled back across heterogeneous ledgers. Instead, failures after irreversible execution are handled by acknowledgements and compensating actions. Table~\ref{tab:transitions} lists the evidence required for each nominal transition.

\begin{table}[!htbp]
\centering\small
\caption{Evidence required for each nominal state transition.}
\label{tab:transitions}
\begin{tabularx}{\linewidth}{@{}>{\raggedright\arraybackslash}p{0.36\linewidth}L@{}}
\toprule
\textbf{Transition} & \textbf{Required evidence}\\
\midrule
\code{CREATED} $\to$ \code{SOURCE\_FINAL} & Source finality evidence satisfying lane policy.\\
\code{SOURCE\_FINAL} $\to$ \code{COMMITTED} & Inclusion of the message commitment in a recognized commitment batch/root.\\
\code{COMMITTED} $\to$ \code{VERIFIED} & Successful lane verifier result.\\
\code{VERIFIED} $\to$ \code{EXECUTING} & Executor authorization and destination submission.\\
\code{EXECUTING} $\to$ \code{SETTLED} & Destination execution receipt/finality satisfying destination policy.\\
\code{SETTLED} $\to$ \code{ACKNOWLEDGED} & Acknowledgement commitment returned or recorded.\\
\bottomrule
\end{tabularx}
\end{table}

%=====================================================================
\section{Commitment Structure}
%=====================================================================

\XIM{} uses commitment roots to summarize message state without replicating complete foreign-chain data. The baseline structure is a Sparse Merkle Tree (SMT)~\cite{merkle1987,dahlberg2016smt} keyed by \code{message\_id}. Implementations MAY use a binary Merkle tree, Verkle commitment~\cite{kuszmaul2018verkle}, or other authenticated dictionary if they preserve equivalent inclusion/non-inclusion and update-verification properties and are identified by \code{commitment\_type}.
\begin{align}
\mathsf{key} &= \mathsf{message\_id}\\
\mathsf{value} &= H\big(\mathsf{state} \,\|\, \mathsf{source\_finality\_commitment} \,\|\, \mathsf{verification\_result} \nonumber\\
&\qquad\;\|\, \mathsf{destination\_tx\_commitment} \,\|\, \mathsf{acknowledgement\_commitment}\big)\\
\mathsf{leaf} &= H(\mathsf{key} \,\|\, \mathsf{value})
\end{align}
Batch roots can be checkpointed to XDC or another coordination ledger. Checkpointing to XDC provides a common audit and dispute anchor but is not required for the correctness of every possible \XIM{} deployment.

%=====================================================================
\section{Append-only Transition Log}
%=====================================================================

\par\noindent\begin{minipage}{\linewidth}
\begin{lstlisting}
XIMEvent {
  message_id
  sequence
  event_type
  network_id
  ledger_position
  timestamp
  previous_state_hash
  new_state_hash
  evidence_hash
  actor_id
}
\end{lstlisting}
\end{minipage}\par\smallskip

For each message, sequence numbers MUST be strictly increasing and contiguous (invariant I6 in Section~\ref{sec:invariants}). The transition log enables deterministic reconstruction of the \XIM{} state, institutional audit trails, incident analysis, and reconciliation. Sensitive source documents SHOULD remain off-chain and be referenced only by cryptographic commitments where appropriate.

%=====================================================================
\section{Verification Framework}
%=====================================================================

Verification is the central security boundary. Each lane binds to a \code{VerificationPolicy} object. No message can be executed solely because a Router asserts that it exists.

\par\noindent\begin{minipage}{\linewidth}
\begin{lstlisting}
VerificationPolicy {
  policy_id
  source_network
  destination_network
  proof_type
  min_finality
  verifier_set_or_contract
  threshold
  max_message_value
  rate_limit
  challenge_period
  policy_version
}
\end{lstlisting}
\end{minipage}\par\smallskip

Table~\ref{tab:proofmodes} lists the supported proof modes. Verification policies SHOULD be value-aware: a low-value informational message may accept a different policy from a high-value asset transfer. Governance MUST define how policies are upgraded and how in-flight messages are treated during upgrades.

\begin{table}[!htbp]
\centering\small
\caption{Verification modes and their trust bases.}
\label{tab:proofmodes}
\begin{tabularx}{\linewidth}{@{}>{\raggedright\arraybackslash}p{0.3\linewidth}LL@{}}
\toprule
\textbf{Proof mode} & \textbf{Typical use} & \textbf{Trust basis}\\
\midrule
\code{NATIVE\_PROOF} & Networks with destination-verifiable state/finality proofs & Source protocol cryptography.\\
\code{LIGHT\_CLIENT} & Consensus-verifiable chains & Correct light-client implementation and source consensus.\\
\code{ZK\_PROOF} & Succinct verification of source state/execution & Proof-system soundness, circuit correctness, source consensus inputs.\\
\code{THRESHOLD\_ATTESTATION} & Networks without practical on-chain light clients & Honest threshold, key security, and governance.\\
\code{TEE\_ATTESTATION} & Specialized gateways & Hardware and attestation trust assumptions.\\
\code{HYBRID} & High-value institutional lanes & A combination, e.g.\ a light-client proof plus an independent risk quorum.\\
\bottomrule
\end{tabularx}
\end{table}

%=====================================================================
\section{Adapter Interface}
%=====================================================================

\par\noindent\begin{minipage}{\linewidth}
\begin{lstlisting}
interface IXIMAdapter {
  networkID()                 -> bytes32
  observe(filter, cursor)     -> Event[]
  finality(event)             -> FinalityEvidence
  buildProof(event)           -> Proof
  verifyProof(message, proof) -> bool
  buildExecution(message)     -> DestinationTransaction
  submit(transaction)         -> TxID
  executionStatus(txID)       -> ExecutionReceipt
  estimateFee(message)        -> FeeQuote
}
\end{lstlisting}
\end{minipage}\par\smallskip

Adapters MUST NOT silently reinterpret \XIM{} fields. Network-specific mappings SHOULD be explicit, versioned, and testable. An adapter upgrade that changes address interpretation, finality, or asset semantics MUST create a new adapter version and SHOULD require lane governance approval.

%=====================================================================
\section{Routing}
%=====================================================================

Direct lanes are sufficient for basic interoperability. \XIM{} additionally defines an optional route-selection layer modeled as a directed graph $G = (V, E)$, where $V$ represents networks or settlement domains and $E$ represents enabled \XIM{} lanes or financial conversion edges. For an edge $e$ and settlement intent $\iota$,
\begin{equation}
\begin{split}
\mathsf{EdgeWeight}(e, \iota) ={}& w_{\mathrm{fee}}\cdot\mathsf{fee}(e) + w_{\mathrm{lat}}\cdot\mathsf{latency}(e) + w_{\mathrm{risk}}\cdot\mathsf{risk}(e)\\
&+ w_{\mathrm{liq}}\cdot\mathsf{liquidity\_penalty}(e) + \mathsf{policy\_penalties}(e, \iota),
\end{split}
\end{equation}
where fee and latency are normalized to a common scale.

Routing MUST be constrained before it is optimized. An edge that fails a hard requirement for jurisdiction, asset support, privacy, verification strength, maximum value, or deadline is removed from the candidate graph. Only then may the router optimize cost or latency. This prevents a cheaper but policy-ineligible path from being selected.

Routers are advisory. A destination verifier MUST independently enforce lane and message policy even if a Router proposes a route.

%=====================================================================
\section{Intent Layer}
%=====================================================================

\par\noindent\begin{minipage}{\linewidth}
\begin{lstlisting}
SettlementIntent {
  intent_id
  payer
  beneficiary
  source_asset
  destination_asset
  amount
  max_fee
  deadline
  required_finality
  privacy_constraints
  jurisdiction_constraints
  compliance_policy_hash
}
\end{lstlisting}
\end{minipage}\par\smallskip

An intent describes the desired economic outcome without binding the requester to a particular network path. One intent may compile into one or more \XIM{} messages. The intent layer is optional and does not alter the security validity of the underlying messages.

%=====================================================================
\section{Stablecoin and Tokenized-Asset Semantics}
%=====================================================================

\XIM{} distinguishes canonical issuer-mediated movement from bridge-issued representations. Where an issuer provides native burn/mint or equivalent cross-domain issuance~\cite{cctp}, \XIM{} SHOULD prefer that mechanism over lock-and-mint wrapping because it avoids accumulating bridge custody risk. The proof policy for such a lane may incorporate issuer attestations in addition to source-chain evidence.

Tokenized securities, deposits, trade-finance instruments, and real-world assets may require transfer restrictions. \XIM{} does not encode legal compliance into a universal global rule; instead, \code{compliance\_policy\_hash} commits to the applicable policy, and destination execution MUST enforce the rules required by the asset representation and participating institutions.

%=====================================================================
\section{Privacy Model}
%=====================================================================

\XIM{} is commitment-oriented and does not require public disclosure of full business payloads. A message may carry \code{payload\_hash}, \code{compliance\_metadata\_hash}, and \code{privacy\_policy\_hash} while encrypted or selectively disclosed data moves through authorized channels. This design is compatible with privacy-preserving institutional systems in which only entitled participants receive transaction details.

Privacy is not provided merely by hashing low-entropy data. Sensitive metadata MUST be encrypted, salted where appropriate, or represented by commitments that resist dictionary attacks. Implementations SHOULD minimize public metadata and SHOULD consider traffic-analysis leakage at the routing layer.

%=====================================================================
\section{ISO 20022 and Legacy Financial Gateways}
%=====================================================================

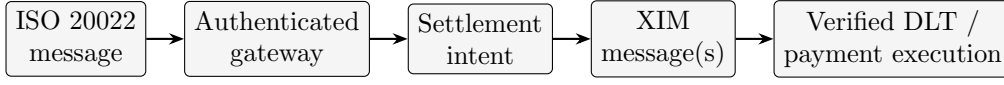
\begin{figure}[!htbp]
\centering
\begin{tikzpicture}[node distance=5mm, every node/.style={font=\scriptsize}]
  \node[bx,minimum width=18mm] (a) {ISO 20022\\message};
  \node[bx,minimum width=18mm,right=of a] (b) {Authenticated\\gateway};
  \node[bx,minimum width=18mm,right=of b] (c) {Settlement\\intent};
  \node[bx,minimum width=18mm,right=of c] (d) {\XIM{}\\message(s)};
  \node[bx,minimum width=18mm,right=of d] (e) {Verified DLT /\\payment execution};
  \draw[arr] (a)--(b); \draw[arr] (b)--(c); \draw[arr] (c)--(d); \draw[arr] (d)--(e);
\end{tikzpicture}
\caption{Mapping a legacy financial instruction into \XIM{}. The gateway is an explicit trust boundary.}
\label{fig:iso}
\end{figure}

\XIM{} gateways may translate authenticated financial instructions into \XIM{} intents and messages (Figure~\ref{fig:iso}). For example, an ISO~20022~\cite{iso20022} \code{pacs.008} instruction can be mapped into a payment intent while preserving a cryptographic reference to the original message. The gateway is a trust boundary: \XIM{} can authenticate the gateway's assertion but cannot independently prove facts that exist only inside a bank's private system unless the bank exposes verifiable evidence.

Gateway deployments SHOULD use HSM-backed signing keys, explicit institution identifiers, key rotation, message deduplication, sanctions and compliance controls where legally required, and auditable mapping rules.

%=====================================================================
\section{XDC Coordination and Settlement Role}
%=====================================================================

XDC Network, whose XDPoS~2.0 consensus provides deterministic finality through a HotStuff-based BFT protocol~\cite{wang2021xdpos}, can act as a neutral checkpoint, coordination, fee, collateral, and settlement domain for \XIM{} deployments. However, \XIM{}'s interoperability semantics do not require every message to transit XDC. This distinction prevents unnecessary token hops while allowing XDC to be used where its execution cost, finality, liquidity, and institutional integrations make it an efficient settlement domain.

{\sloppy Potential XDC-native components include \code{XIMRouter}, \code{XIMMessageRegistry}, \code{XIMVerifierRegistry}, \code{XIMAssetRegistry}, \code{XIMAdapterRegistry}, and \code{XIMRiskManager}. Optional modules include liquidity, fee, and compliance registries.\par}

%=====================================================================
\section{Reference Smart-Contract Interfaces}
%=====================================================================

\par\noindent\begin{minipage}{\linewidth}
\begin{lstlisting}[language=Java,morekeywords={function,external,view,returns,bytes32,bytes,uint8,uint32,uint256,address,calldata,interface}]
interface IXIMMessageRegistry {
  function commit(bytes32 messageId, bytes32 messageCommitment) external;
  function markVerified(bytes32 messageId, bytes32 proofHash) external;
  function markSettled(bytes32 messageId, bytes32 destinationTxHash) external;
  function acknowledge(bytes32 messageId, bytes32 ackHash) external;
  function status(bytes32 messageId) external view returns (uint8);
}

interface IXIMVerifierRegistry {
  function verifier(bytes32 laneId, uint32 policyVersion)
      external view returns (address);
  function verify(bytes32 laneId, bytes calldata message, bytes calldata proof)
      external view returns (bool);
}

interface IXIMRiskManager {
  function laneStatus(bytes32 laneId) external view returns (uint8);
  function valueLimit(bytes32 laneId) external view returns (uint256);
}
\end{lstlisting}
\end{minipage}\par\smallskip

%=====================================================================
\section{Security Properties}
%=====================================================================

Subject to the explicit assumptions of the selected lane verification policy, \XIM{} aims to provide the properties in Table~\ref{tab:props}.

\begin{table}[!htbp]
\centering\small
\caption{Target security properties (informal).}
\label{tab:props}
\begin{tabularx}{\linewidth}{@{}>{\raggedright\arraybackslash}p{0.25\linewidth}L@{}}
\toprule
\textbf{Property} & \textbf{Informal requirement}\\
\midrule
Authenticity & A destination accepts a source assertion only if the required verification evidence validates.\\
Integrity & Any mutation of a canonical message changes its \code{message\_id} and commitment.\\
Replay resistance & A consumed \code{message\_id} cannot execute twice on the same destination security domain.\\
Ordering where required & Per-channel or application sequence constraints can be enforced.\\
Timeout safety & Expired messages cannot newly execute unless the application explicitly permits late settlement.\\
Auditability & State transitions are reconstructible from authenticated commitments and event records.\\
Lane isolation & A compromise or anomaly on one lane can be paused without globally halting unrelated lanes.\\
Policy explicitness & Security assumptions are bound to versioned lane policies rather than hidden in relayer software.\\
\bottomrule
\end{tabularx}
\end{table}

%=====================================================================
\section{Threat Model}
%=====================================================================

Table~\ref{tab:threats} summarizes the principal threats, their mitigations, and residual limitations.

\begin{table}[!htbp]
\centering\small
\caption{Threats, mitigations, and limitations.}
\label{tab:threats}
\begin{tabularx}{\linewidth}{@{}>{\raggedright\arraybackslash}p{0.3\linewidth}L@{}}
\toprule
\textbf{Threat} & \textbf{Mitigation / limitation}\\
\midrule
Malicious router & Router output is advisory; the destination verifies proof and policy independently.\\
Compromised executor & The executor cannot create valid source evidence; the destination contract enforces verification and replay protection.\\
Source-chain reorganization & Minimum finality policy; probabilistic-finality safety margins; delayed high-value execution.\\
Verifier quorum compromise & Threshold diversification, stake/bonding where appropriate, key rotation, independent risk monitor, lane value limits.\\
Smart-contract bug & Small modular contracts, audits, formal verification of critical state transitions, staged limits.\\
Replay across chains or protocol versions & Domain-separated message hash including source/destination identifiers and protocol version.\\
Proof-system vulnerability & Versioned verifier registry, emergency lane pause, multiple proof modes or hybrid verification.\\
Gateway key compromise & HSMs, rate/value limits, multi-party authorization, anomaly detection, rapid key revocation.\\
Liquidity manipulation & Hard slippage/liquidity constraints and independent quote validation.\\
Privacy leakage & Commitment-only public fields, encryption, metadata minimization; traffic analysis remains a residual risk.\\
\bottomrule
\end{tabularx}
\end{table}

\subsection{Safety versus Liveness}
\XIM{} prioritizes safety for asset-moving lanes. If finality, verifier availability, or risk status is uncertain, the protocol SHOULD halt or delay execution rather than guess. Liveness therefore depends on source availability, verifier availability, destination availability, and governance responsiveness.

%=====================================================================
\section{Atomicity and Multi-leg Settlement}
\label{sec:atomicity}
%=====================================================================

True atomicity across unrelated consensus systems is not always achievable without shared locking or a common synchronizer~\cite{zamyatin2021sok}. \XIM{} therefore distinguishes (i)~native atomic execution within one domain, (ii)~conditional cross-domain settlement using escrow, hash, or time conditions or application locks~\cite{herlihy2018atomic}, and (iii)~coordinated but non-atomic settlement with compensating actions. Implementations MUST accurately advertise which model applies.

For institutional delivery-versus-payment, a deployment may place both legs under a common conditional settlement contract or domain, or use synchronized application-level locks with bounded timeouts. \XIM{} provides message coordination but does not claim that messaging alone eliminates principal risk.

%=====================================================================
\section{Risk Management and Circuit Breakers}
\label{sec:risk}
%=====================================================================

An independent Risk Monitor SHOULD evaluate source finality anomalies, unusual mint/burn volumes, verifier disagreement, rapid value spikes, proof failures, adapter inconsistencies, and chain incidents~\cite{lee2023bridges}. Risk controls SHOULD be lane-scoped and may include:
\begin{itemize}
  \item per-message value caps;
  \item rolling hourly and daily lane limits;
  \item velocity limits per asset and participant;
  \item emergency pause for new executions while allowing safe acknowledgement and refund paths;
  \item verifier-set diversity requirements;
  \item delayed settlement for unusually large transfers; and
  \item automatic downgrade to stricter verification when risk thresholds are exceeded.
\end{itemize}

%=====================================================================
\section{Governance and Upgradeability}
%=====================================================================

Governance controls verifier policies, adapter versions, asset registry entries, lane activation, limits, and emergency actions. Governance itself is part of the threat model. Production deployments SHOULD use timelocks for non-emergency upgrades, multi-party authorization, transparent change records, and explicit emergency powers with post-event review.

Protocol versioning MUST preserve deterministic interpretation of historical messages. A major version change SHOULD use a new domain separator.

%=====================================================================
\section{Performance Model}
%=====================================================================

End-to-end latency is bounded by the slowest mandatory stage rather than by XDC block time alone:
\begin{equation}
\begin{split}
T_{\mathrm{total}} ={}& T_{\mathrm{source\_finality}} + T_{\mathrm{observation}} + T_{\mathrm{proof\_generation}} + T_{\mathrm{verification}}\\
&+ T_{\mathrm{routing}} + T_{\mathrm{destination\_submission}} + T_{\mathrm{destination\_finality}} + T_{\mathrm{ack}}.
\end{split}
\end{equation}

For threshold-attested lanes, proof generation may be short but carries stronger trust assumptions. For light-client or ZK lanes, cryptographic verification may reduce trust but increase engineering complexity or proof latency. \XIM{} makes this trade-off explicit at the lane-policy level.

Throughput can be improved through message batching. If $k$ messages are committed under one Merkle root, checkpoint cost is amortized across $k$ messages. Executors MAY batch destination submissions where destination semantics permit.

%=====================================================================
\section{Reference Node Architecture}
%=====================================================================

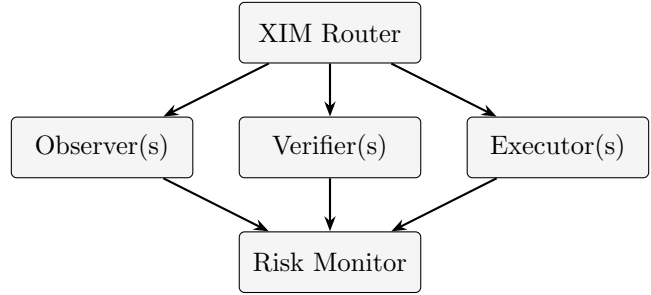
\begin{figure}[!htbp]
\centering
\begin{tikzpicture}[node distance=7mm and 6mm]
  \node[bx] (obs) {Observer(s)};
  \node[bx,right=of obs] (ver) {Verifier(s)};
  \node[bx,right=of ver] (exe) {Executor(s)};
  \node[bx,above=of ver] (rt) {\XIM{} Router};
  \node[bx,below=of ver] (rm) {Risk Monitor};
  \draw[arr] (rt) -- (obs); \draw[arr] (rt) -- (ver); \draw[arr] (rt) -- (exe);
  \draw[arr] (obs) -- (rm); \draw[arr] (ver) -- (rm); \draw[arr] (exe) -- (rm);
\end{tikzpicture}
\caption{Reference node architecture. The Risk Monitor observes all execution components independently of the Router.}
\label{fig:node}
\end{figure}

Figure~\ref{fig:node} shows the reference node architecture. A reference implementation can use PostgreSQL for durable message and application metadata, Redis for ephemeral route and quote caches, RocksDB for authenticated state, and a graph engine or in-memory adjacency representation for routing. The cryptographic commitment root, not the database engine, defines protocol state integrity.

%=====================================================================
\section{Illustrative Workflows}
%=====================================================================

\subsection{Ethereum to XDC Asset Transfer}
\begin{enumerate}
  \item A user submits a transfer intent specifying source asset, XDC destination, amount, maximum fee, and verification requirement.
  \item The Ethereum adapter observes the source event after the required finality threshold.
  \item The Observer constructs a canonical \code{XIMMessage} and deterministic \code{message\_id}.
  \item The proof builder generates the lane-required evidence (e.g., a light-client or state proof, a ZK proof, or an approved attestation).
  \item The message commitment enters a batch and is checkpointed according to deployment policy.
  \item The XDC verifier contract validates the evidence and checks lane status, expiry, limits, and replay state.
  \item The Executor calls the destination asset or settlement contract.
  \item The destination receipt is recorded as \code{SETTLED}.
  \item An acknowledgement is generated and optionally returned to the source application.
\end{enumerate}

\subsection{Canton-connected Institutional Workflow}
A Canton-connected gateway or adapter observes an authorized event available to the relevant participant. It constructs a \XIM{} commitment without publishing confidential transaction contents. Depending on available Canton integration primitives, the lane may verify a cryptographic proof, an authorized participant or synchronizer attestation, or a hybrid policy. \XIM{} then routes the verified instruction to XDC or another destination. Sensitive documents and participant-specific details remain within entitled systems; public \XIM{} state contains only the minimum commitments needed for integrity, settlement, and audit.

This workflow deliberately avoids assuming that Canton exposes Ethereum-like public state proofs. The concrete adapter and proof policy must follow the capabilities and security model of the deployed Canton integration~\cite{canton-protocol,canton2020}.

\subsection{ISO 20022 to XDC Settlement}
\begin{enumerate}
  \item An institution receives or originates an ISO~20022 payment instruction.
  \item An authenticated \XIM{} Gateway validates schema, authorization, duplication status, and institution policy.
  \item The gateway creates a \code{SettlementIntent} and commits to the source financial message hash.
  \item The policy engine selects an eligible settlement route, for example native USDC on XDC or another approved rail.
  \item Gateway signature and HSM evidence is verified under the bank-to-\XIM{} lane policy.
  \item Destination execution occurs and the resulting transaction identifier is bound to the \XIM{} message.
  \item Reconciliation output maps \XIM{} settlement status back to the institution's payment workflow.
\end{enumerate}

%=====================================================================
\section{Formal Invariants}
\label{sec:invariants}
%=====================================================================

A conforming \XIM{} implementation SHOULD preserve at least the following invariants, where $m$ denotes a message, $d$ a destination security domain, and $\prec$ the lifecycle order of Figure~\ref{fig:lifecycle}.

\begin{description}[leftmargin=1.2cm,labelwidth=1cm,font=\normalfont\bfseries]
  \item[I1] \emph{At-most-once destination execution:} $\;\mathsf{executed}[d, m.\mathsf{message\_id}] \le 1$.
  \item[I2] \emph{Verification before execution:} $\;\mathsf{EXECUTING}(m) \Rightarrow \mathsf{VERIFIED}(m)$.
  \item[I3] \emph{Expiry:} $\;\mathsf{now} > m.\mathsf{expiry} \,\wedge\, \mathsf{state}(m) \prec \mathsf{EXECUTING} \Rightarrow$ no new execution of $m$.
  \item[I4] \emph{Commitment integrity:} $\;\mathsf{accepted}(m) \Rightarrow H(\mathsf{domain\_separator}\,\|\,\mathsf{canonical\_encode}(m)) = m.\mathsf{message\_id}$.
  \item[I5] \emph{Lane authorization:} $\;\mathsf{execution}(m) \Rightarrow \mathsf{active}(\mathsf{lane\_id}, \mathsf{policy\_version})$.
  \item[I6] \emph{Monotonic sequence:} $\;\mathsf{event}[n+1].\mathsf{sequence} = \mathsf{event}[n].\mathsf{sequence} + 1$.
  \item[I7] \emph{Asset mapping:} $\;\mathsf{transfer}(m, r) \Rightarrow r$ is authorized for $m.\mathsf{asset\_id}$ on the destination.
\end{description}

%=====================================================================
\section{Implementation Plan and Research Roadmap}
%=====================================================================

\subsection{Staged Minimum Viable Implementation}
Table~\ref{tab:mvp} outlines a staged twelve-week plan for a minimum viable implementation. No performance or security claims in this paper are yet backed by measurements from this implementation; the pilot phase is intended to produce a reproducible benchmark report.

\begin{table}[!htbp]
\centering\small
\caption{Staged twelve-week implementation plan.}
\label{tab:mvp}
\begin{tabularx}{\linewidth}{@{}>{\raggedright\arraybackslash}p{0.22\linewidth}>{\raggedright\arraybackslash}p{0.1\linewidth}L@{}}
\toprule
\textbf{Phase} & \textbf{Weeks} & \textbf{Deliverables}\\
\midrule
Protocol freeze & 1--2 & \XIM{} v0.1 schema; NetworkID/UAID format; state machine; lane policy model; test vectors.\\
XDC contracts & 2--5 & MessageRegistry, VerifierRegistry, RiskManager, replay protection, event model.\\
First adapters & 3--7 & XDC and Ethereum observers/executors; deterministic encoding; threshold-attestation verifier as bootstrap.\\
Commitments and SDK & 5--8 & SMT library, batch roots, TypeScript/Go SDKs, REST/gRPC API.\\
Security and operations & 7--10 & Rate limits, lane pause, key rotation, monitoring, chaos/failure tests.\\
Pilot & 10--12 & Testnet Ethereum$\leftrightarrow$XDC message transfer and one asset-transfer demonstration; benchmark report.\\
\bottomrule
\end{tabularx}
\end{table}

\subsection{Research Roadmap}
\begin{itemize}
  \item Replace or supplement bootstrap attestations with light-client or ZK verification on high-value public-chain lanes.
  \item Formal specification in TLA+~\cite{lamport2002tla} or an equivalent formalism for message lifecycle, replay resistance, timeout, and acknowledgement invariants.
  \item Formal verification of critical Solidity contracts.
  \item A Canton adapter feasibility study with an explicit privacy and proof model.
  \item An ISO~20022 gateway reference mapping and institutional HSM profile.
  \item A multi-route liquidity and policy engine.
  \item Cross-domain conditional settlement and delivery-versus-payment research.
  \item Verkle or alternative vector commitments if proof-size or update characteristics justify migration from SMTs.
\end{itemize}

%=====================================================================
\section{Security Limitations}
%=====================================================================

\XIM{} cannot make a compromised source ledger truthful. If a source consensus finalizes an invalid economic state according to its own governance, a correct light client will faithfully prove that state. Similarly, attestation-based gateways inherit signer and key-management risk. Cross-chain systems also inherit software risk from adapters, verifiers, destination contracts, and asset issuers. \XIM{}'s objective is therefore not to eliminate trust, but to make trust assumptions explicit, modular, measurable, and isolatable.

Multi-hop routes compound risk. A route is no stronger than the weakest economically relevant edge unless the application uses additional end-to-end verification. High-value routes SHOULD minimize trust-domain transitions.

%=====================================================================
\section{Discussion}
%=====================================================================

\XIM{} defines interoperability around canonical authenticated messages rather than a shared execution environment. This allows heterogeneous networks to retain their native consensus, privacy, governance, and execution models while participating in a common message lifecycle and verification framework. Non-blockchain systems may also participate through explicitly trusted and auditable gateways.

Within the XDC ecosystem, XDC Subnets provide application-specific or permissioned execution environments with configurable governance and privacy. Connectivity between a subnet and the XDC mainnet can be provided through XDC Zero, which supports cross-domain message and asset transfer using relayers, validation components, endpoints, and mainnet checkpointing.

XDC may therefore serve as an optional coordination, checkpointing, and settlement domain for \XIM{} deployments without being a mandatory routing layer. This separation preserves protocol neutrality while allowing applications to use XDC where its settlement characteristics, infrastructure, or ecosystem integrations satisfy the requirements of a particular workflow.

%=====================================================================
\section{Conclusion}
%=====================================================================

\XIM{} proposes a modular interledger messaging protocol for a financial world composed of many independent state machines. The protocol separates canonical message semantics, cryptographic commitments, verification, routing, execution, asset identity, privacy metadata, and risk controls. It supports a spectrum of verification models instead of assuming that one bridge-security model fits every network. By making security assumptions explicit per lane and by minimizing foreign-state replication, \XIM{} aims to provide a practical foundation for public-chain interoperability, institutional tokenized assets, stablecoin settlement, privacy-preserving ledgers, and authenticated connections to traditional financial rails.

The next stage of the work is implementation and measurement. Claims about throughput, latency, security, and operational resilience should be validated through public test vectors, reproducible benchmarks, adversarial testing, audits, and formal analysis before production use.

%=====================================================================

%=====================================================================
\appendix
%=====================================================================

\section{Proposed \XIM{} Protobuf Skeleton}
\label{app:proto}

\par\noindent\begin{minipage}{\linewidth}
\begin{lstlisting}
syntax = "proto3";
package xim.v1;

message XIMMessage {
  uint32 version                  = 1;
  bytes  source_network           = 2;
  uint64 source_height            = 3;
  bytes  source_tx_id             = 4;
  bytes  sender                   = 5;
  bytes  destination_network      = 6;
  bytes  receiver                 = 7;
  uint32 action_type              = 8;
  bytes  asset_id                 = 9;
  bytes  amount                   = 10;
  bytes  payload_hash             = 11;
  uint64 nonce                    = 12;
  uint64 created_at               = 13;
  uint64 expiry                   = 14;
  uint32 finality_policy          = 15;
  uint32 proof_type               = 16;
  bytes  proof_commitment         = 17;
  bytes  compliance_metadata_hash = 18;
  bytes  privacy_policy_hash      = 19;
  bytes  route_policy_hash        = 20;
}
\end{lstlisting}
\end{minipage}\par\smallskip

\section{API Sketch}

\par\noindent\begin{minipage}{\linewidth}
\begin{lstlisting}
POST /v1/intents
{
  "source_network": "ethereum",
  "destination_network": "xdc",
  "source_asset": "UAID:...",
  "destination_asset": "UAID:...",
  "amount": "10000000",
  "beneficiary": "xdc:...",
  "constraints": {
    "max_fee": "500",
    "deadline": 1780000000,
    "verification_class": "institutional"
  }
}

GET /v1/messages/{message_id}
GET /v1/routes/{intent_id}
GET /v1/proofs/{message_id}
GET /v1/assets/{uaid}
\end{lstlisting}
\end{minipage}\par\smallskip

\section{Test Vector Requirements}
A conforming implementation SHOULD publish test vectors covering:
\begin{itemize}
  \item the canonical encoding byte sequence for each protocol version;
  \item the expected \code{message\_id} for each canonical message;
  \item Merkle leaf, path, and root vectors;
  \item replay-rejection tests;
  \item expiry boundary tests;
  \item policy-version upgrade tests;
  \item source reorganization and finality tests;
  \item verifier disagreement and lane-pause tests;
  \item destination execution failure and compensation tests; and
  \item cross-language SDK conformance tests.
\end{itemize}

\end{document}